\documentclass{menu99}    
\usepackage{epsfig}

\begin{document}
    \setlength{\baselineskip}{2.6ex}

\title{New Result for the  Pion--Nucleon Sigma Term from an Updated 
VPI/GW $\pi$N Partial--Wave and Dispersion Relation Analysis}

\author{
M.M. Pavan \thanks{Present address: TRIUMF, Vancouver,
  B.C. V6T-2A3 ;  EMAIL: marcello.pavan@triumf.ca} \\
  \emph{Lab for Nuclear Science, M.I.T., Cambridge, MA 02139} \\
R.A. Arndt \thanks{Present affiliation: Dept. of Physics, The George
  Washington University,   Washington, D.C. 20052 }\\
  \emph{Dept. of Physics, Virginia Polytechnic and State University,
    Blacksburg, VA 24061} \\
I.I. Strakovsky and R.L. Workman \\
  \emph{Dept. of Physics, The George Washington University, Washington,
    D.C. 20052}
}

\maketitle

\begin{abstract} \setlength{\baselineskip}{2.6ex}
  A new result for the $\pi$N sigma term from an updated $\pi$N
  partial--wave and dispersion relation analysis of the Virginia
  Polytechnic Institute (now George Washington University) group is
  discussed. Using a method similar to that of Gasser, Leutwyler,
  Locher, and Sainio, we obtain $\Sigma=$90$\pm$8 MeV (preliminary), in
  disagreement with the canonical result 64$\pm$8 MeV, but consistent with
  expectations based on new information on the $\pi$NN coupling constant,
  pionic atoms, and the $\Delta$ resonance width.
\end{abstract}

\setlength{\baselineskip}{2.6ex}

\section*{Introduction}

The pion nucleon sigma term ($\Sigma$) continues to be a puzzle some thirty
years after initial attempts to determine it.  The keen interest in
$\Sigma$ comes from the fact that it vanishes in the massless quark
(chiral) limit of QCD, and becomes non-zero only for a non-zero light (up
or down) quark mass, so it is a crucial parameter in the understanding of
chiral symmetry breaking (see e.g.  Refs.~\cite{gl82,glls88}).  The
nucleon's strange quark content can be inferred from $\Sigma$ (see e.g.
Ref.~\cite{glls88}), so $\Sigma$ is also relevant to {\it quark
  confinement}, not yet fully understood, since one must understand the
mechanism for accommodating strange quarks in an ostensibly light quark
object ~\cite{jaffe-pc}. Thus $\Sigma$ is a parameter of {\it fundamental}
significance to low energy QCD, making it crucial to obtain its value as
precisely as possible.  The canonical result for $\Sigma \simeq 64$ MeV
~\cite{koch82,hoehler} implies a large nucleon strangeness content
~\cite{glls88}, and much effort has been spent trying to understand that.
This article outlines recent work of the (former) Virginia Polytechnic
Institute (VPI), (now George Washington University (GWU)) group to extract
the ``experimental'' value of the sigma term ($\Sigma$) from the $\pi$N
scattering data as part of ongoing $\pi$N partial-wave (PWA) and dispersion
relation (DR) analyses.

\section*{Experimental $\Sigma$ Term}

The ``experimental'' sigma term $\Sigma$ is related to the $\pi$N isoscalar
amplitude $\bar{D}^{+}$ (bar signifies the pseudovector Born term is
subtracted) at the ``Cheng-Dashen point'' ~\cite{cheng71}:
 \begin{equation}
   \label{eqn:sigmaDef}
 \Sigma = F^{2}_{\pi}\bar{D}^{+}(\nu=0,t=2m^{2}_{\pi}) 
  \end{equation}
where $F_{\pi}$=92.4 MeV is the pion decay constant, $\nu$ is the
crossing energy variable, and $t$ is the four-momentum transfer.  Since
the Cheng-Dashen point lies outside the physical $\pi$N scattering
region, the experimental $\pi$N amplitudes must be {\it extrapolated} in
order to obtain $\Sigma$.  The most theoretically well-founded
extrapolation approach is based on dispersion relation (DR) analyses of
the scattering amplitudes ~\cite{hoehler}.  In the early 80$^{s}$, the
Karlsruhe-Helsinki group performed extensive investigations into obtaining
$\Sigma$ from $\pi$N dispersion relations ~\cite{hoehler}.  The canonical
result $\Sigma = 64 \pm 8$ MeV was based on hyperbolic dispersion relation
~\cite{koch82} calculations using the groups' $\pi$N ~\cite{kh80} and
$\pi\pi$ ~\cite{hoehler} phase shifts.
 
\begin{figure}[ht]
  {\par\centering
\begin{tabular}{ c c }
  \epsfig{file= 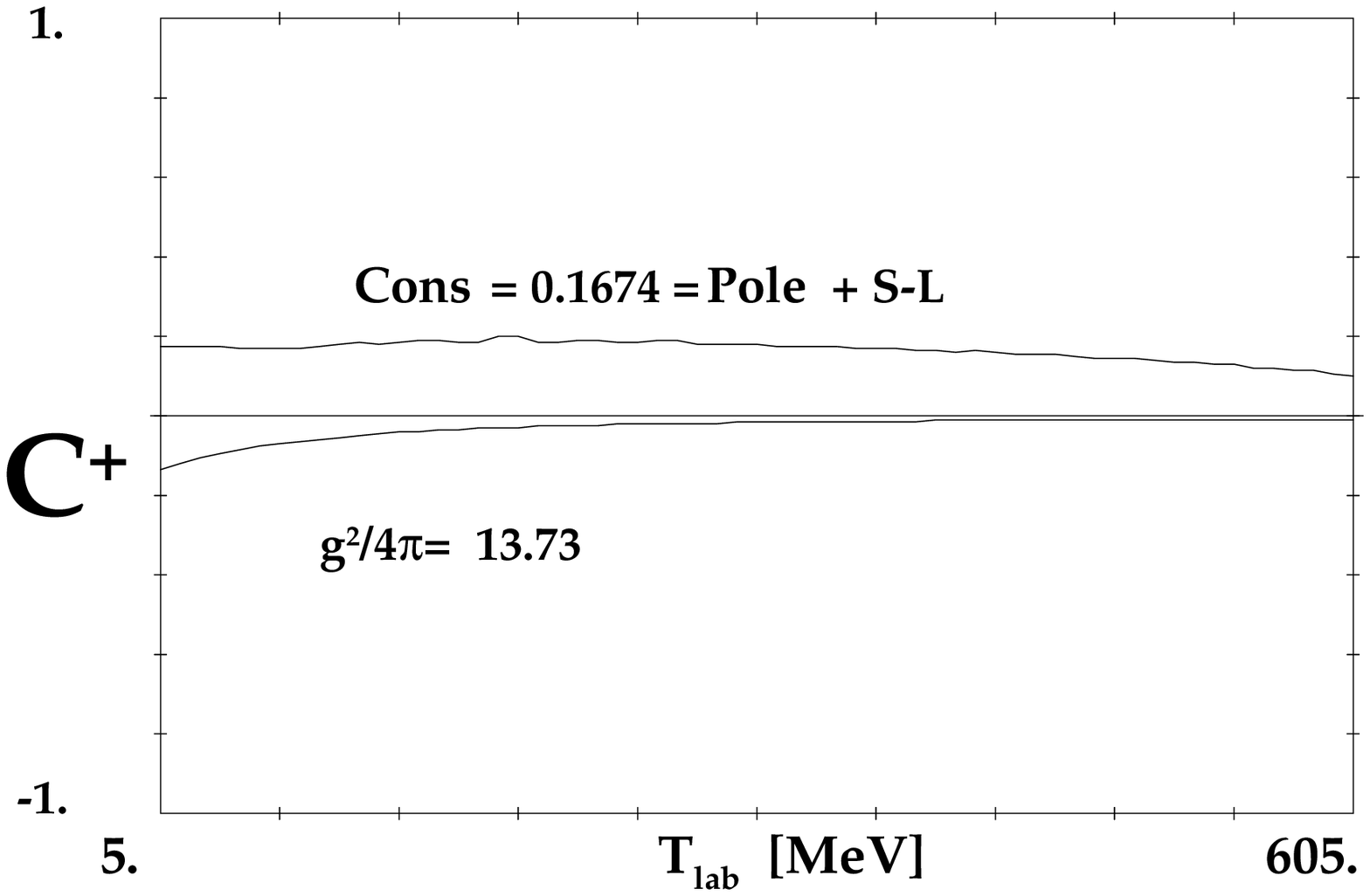, width=0.4\columnwidth} &
  \epsfig{file= 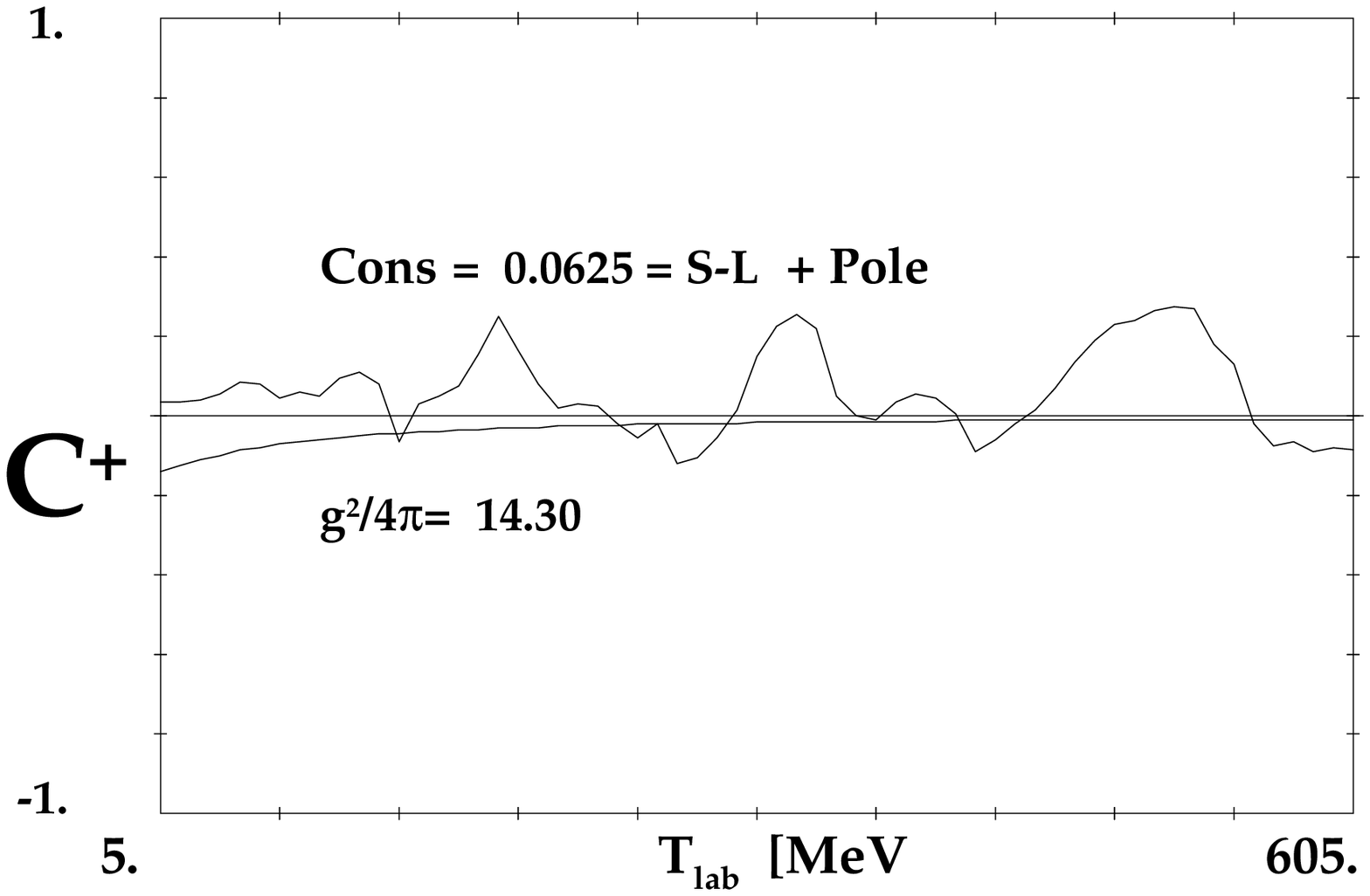, width=0.4\columnwidth}
\end{tabular}
 \par}
  
\caption{\small \setlength{\baselineskip}{2.6ex} \label{fig:c+Const}
{\bf Left:} Subtraction constant (``{\bf s}cattering {\bf l}ength + pole'')
 and  Born term in the forward
$C^+$  dispersion relation as a function of energy from our $\pi$N analysis
SM99 ;  {\bf Right:} same, for the Karlsruhe KA84 analysis
\protect~\cite{ka84}.  This DR yields the coefficient $\bar{d}^{+}_{00}$
in Eqn.\protect~\ref{eqn:linearSigma}.
}
\end{figure}

The only recent dispersion theoretic determinations have been by Sainio
~\cite{sainio97}, based on the method of Gasser, Leutwyler, Locher, and
Sainio (GLLS)~\cite{glls88}. The method exploits the fact that
$\bar{D}^{+}(t)$ can be expressed as a power series in $t$ ~\cite{hoehler},
the coefficients determined from dispersion relation subtraction constants.
The coefficients up to $O(t)$, $\bar{d}^{+}_{00}$ and $\bar{d}^{+}_{01}$,
are determined from the forward $\bar{C}^{+}$ and ``derivative''
$\bar{E}^{+}$ DRs, respectively. The smaller $O(\geq t^{2})$ correction
$\Delta_{D}\simeq$12 MeV is determined employing $\pi\pi N\bar{N}$ phase
shifts (15 MeV), and $\Delta$ isobar exchange (-3 MeV) ~\cite{glls88}.
$\Sigma$ is then expressed as:

 \begin{equation}
   \label{eqn:linearSigma}
\Sigma =F^2_{\pi}\cdot(\bar{d}^+_{00}+2m^2_{\pi}\cdot\bar{d}^+_{01})
                        +\Delta_{D} \equiv \Sigma_{d} + \Delta_{D}  
 \end{equation}

 In the GLLS approach, the Karlsruhe KH80~\cite{kh80} or KA84~\cite{ka84}
 $\pi$N phases shifts are used as {\em fixed input} above about
 T$_{\pi}$=70 MeV, and the D and higher phases are used below the cutoff as
 well in six forward dispersion relations ($B^{\pm}, C^{\pm}, E^{\pm}$). By
 fitting the low energy data, $\bar{d}^{+}_{00}$ and $\bar{d}^{+}_{01}$ can
 be determined.  Their result ~\cite{glls88,sainio97} was
 $\Sigma_{d}\simeq$50 MeV, and $\Delta_{D}$=12 MeV, leading to
 $\Sigma\sim$ 62 MeV, in agreement with the Karlsruhe results
 ~\cite{hoehler,koch82}. However, since the dispersion relations were
 constrained to be satisfied, the subtraction constants, which are energy
 {\em independent}, must be the {\em same} at low energies where the data
 were fit as at high energies where they were {\em fixed} input. Therefore,
 $\Sigma_d$ could not have come out significantly different than the
 Karlsruhe result.  Nonetheless, this analysis provided a very useful {\it
   validation} of the method.  The technique has been criticized
 ~\cite{hoehler-pc} since the $E$ DR is more sensitive to the higher
 partial waves than the other DRs, so it {\it could be} rather uncertain
 due to uncertainty in the higher phases. What the GLLS analyses showed was
 that this is in fact not the case, and the method can be used reliably to
 extract $\Sigma_d$.

\begin{figure}[ht]
  {\par\centering
\begin{tabular}{ c c }
  \epsfig{file= 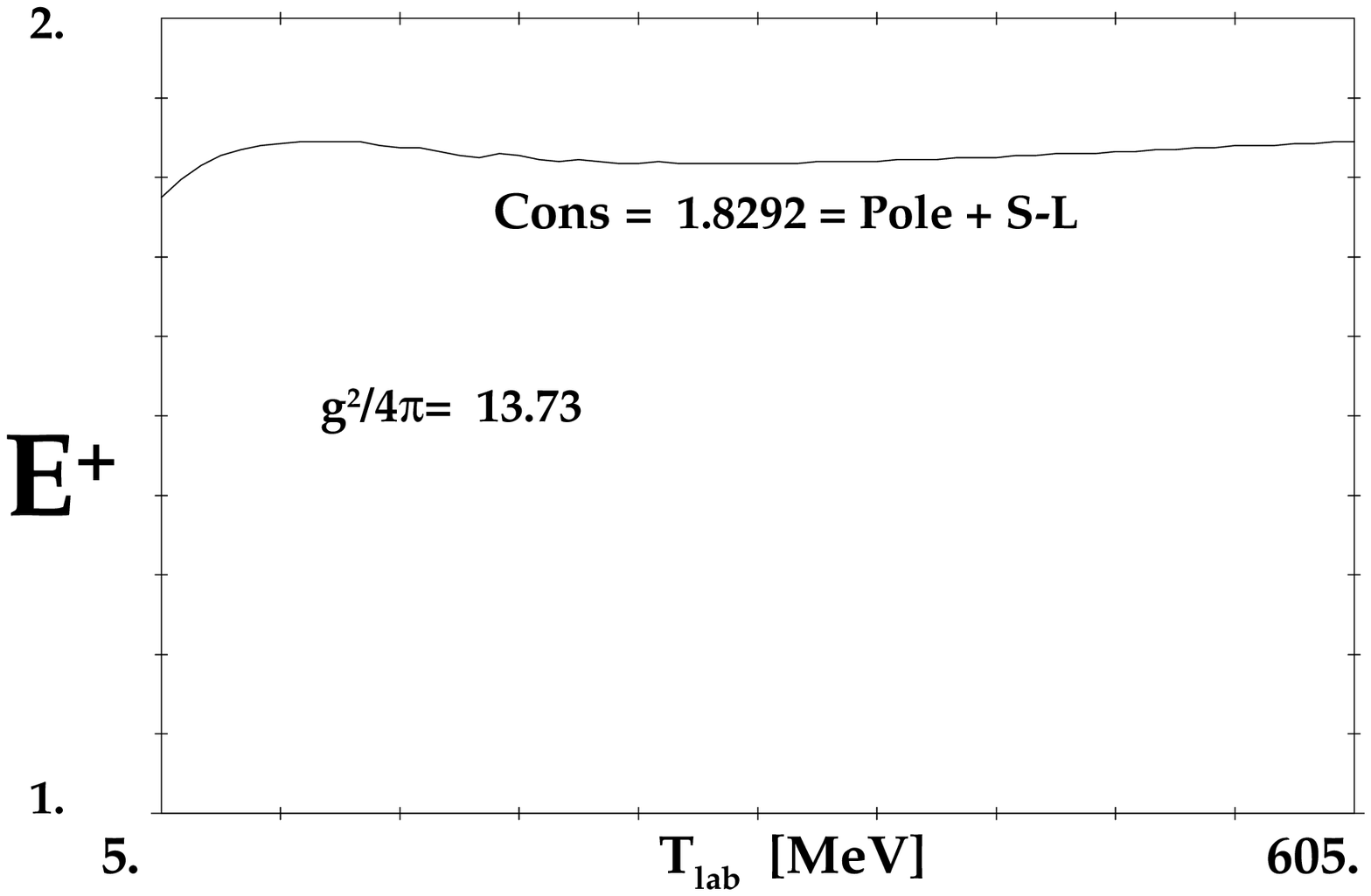, width=0.4\columnwidth} &
  \epsfig{file= 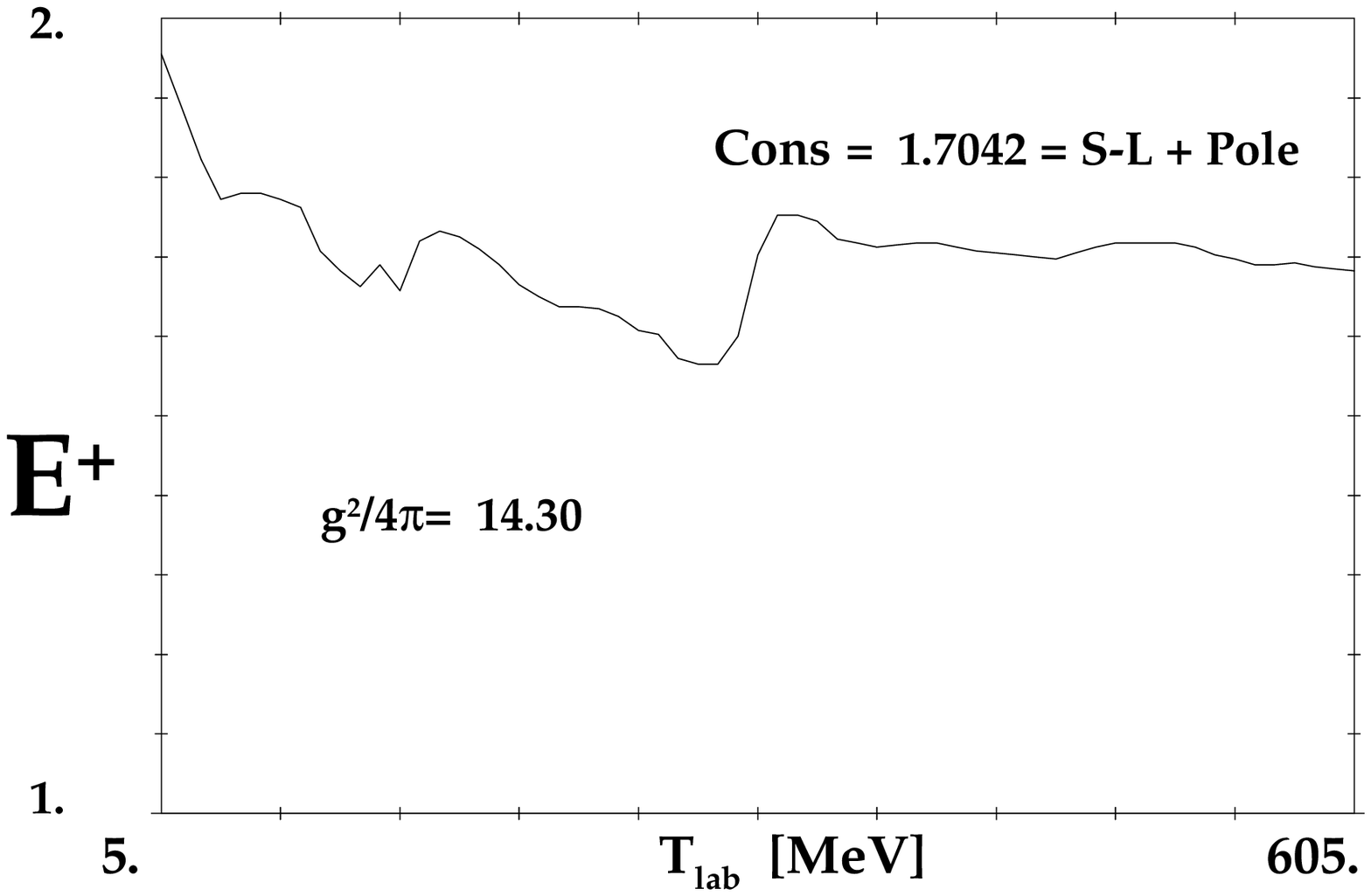, width=0.4\columnwidth}
\end{tabular}
 \par}
  
\caption{\small \setlength{\baselineskip}{2.6ex} \label{fig:e+Const}
{\bf Left:} Subtraction constant in the forward
$E^+$  dispersion relation as a function of energy from our $\pi$N analysis
SM99 ;  {\bf Right:} same, for the Karlsruhe KA84 analysis
\protect~\cite{ka84}. This DR yields the coefficient $\bar{d}^{+}_{01}$
in Eqn.\protect~\ref{eqn:linearSigma}.
}
\end{figure}

Since the GLLS analyses simply demonstrated another method to get
$\Sigma_d$ from the KH80 $\pi$N analysis, there have been {\em no} recent
DR--based Sigma term analyses independent of the results of the Karlsruhe
group ~\cite{hoehler,koch82}.  Consequently, our group has developed a
version of the GLLS technique as part of our own $\pi$N partial-wave and
dispersion relation analysis.  The method will be outlined in the following
sections.

\section*{VPI/GW $\Sigma$ Term Analysis Method}

\begin{figure}
  {\par\centering
\begin{tabular}{ c c }
  \epsfig{file= 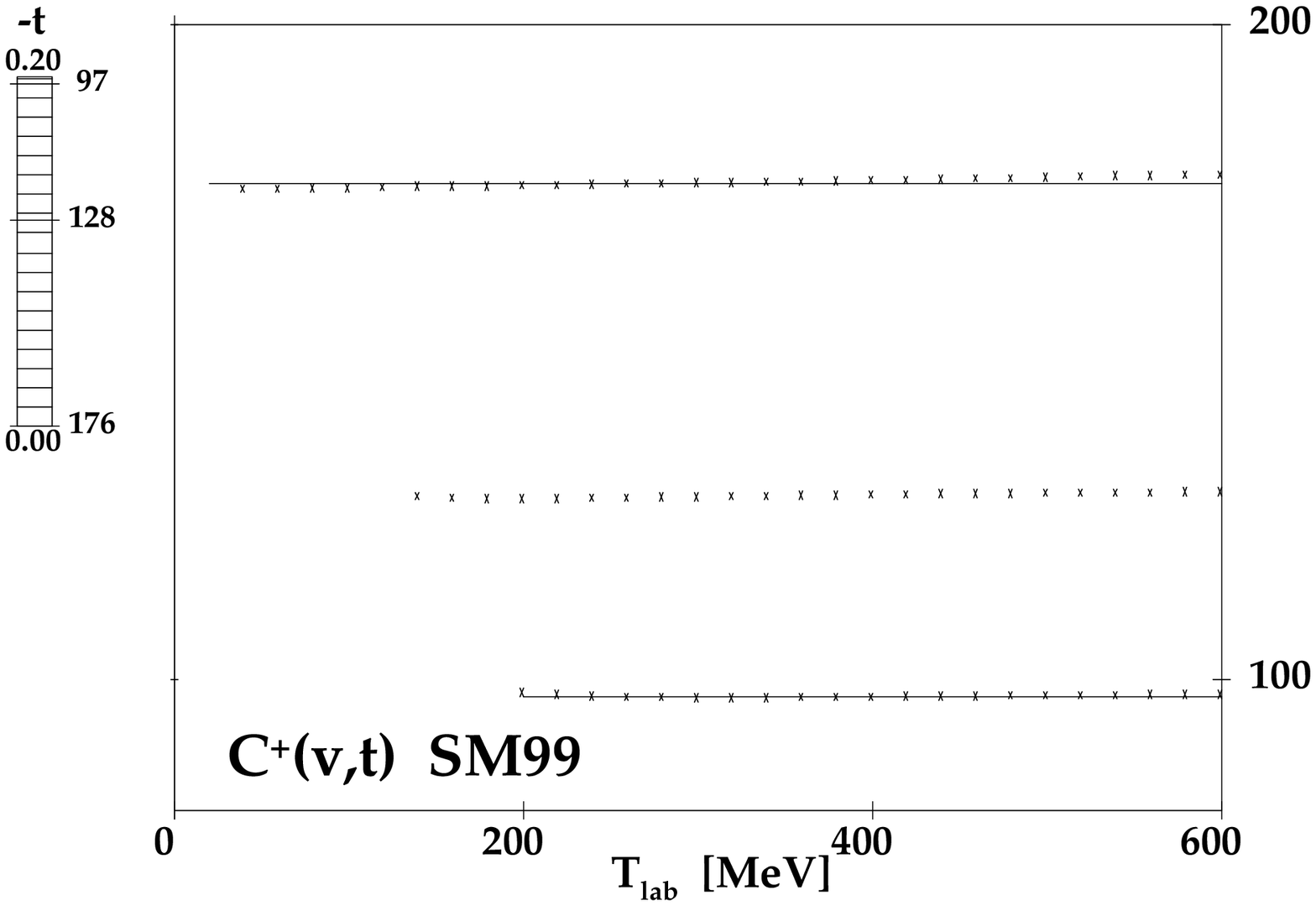, width=0.45\columnwidth} &
  \epsfig{file= 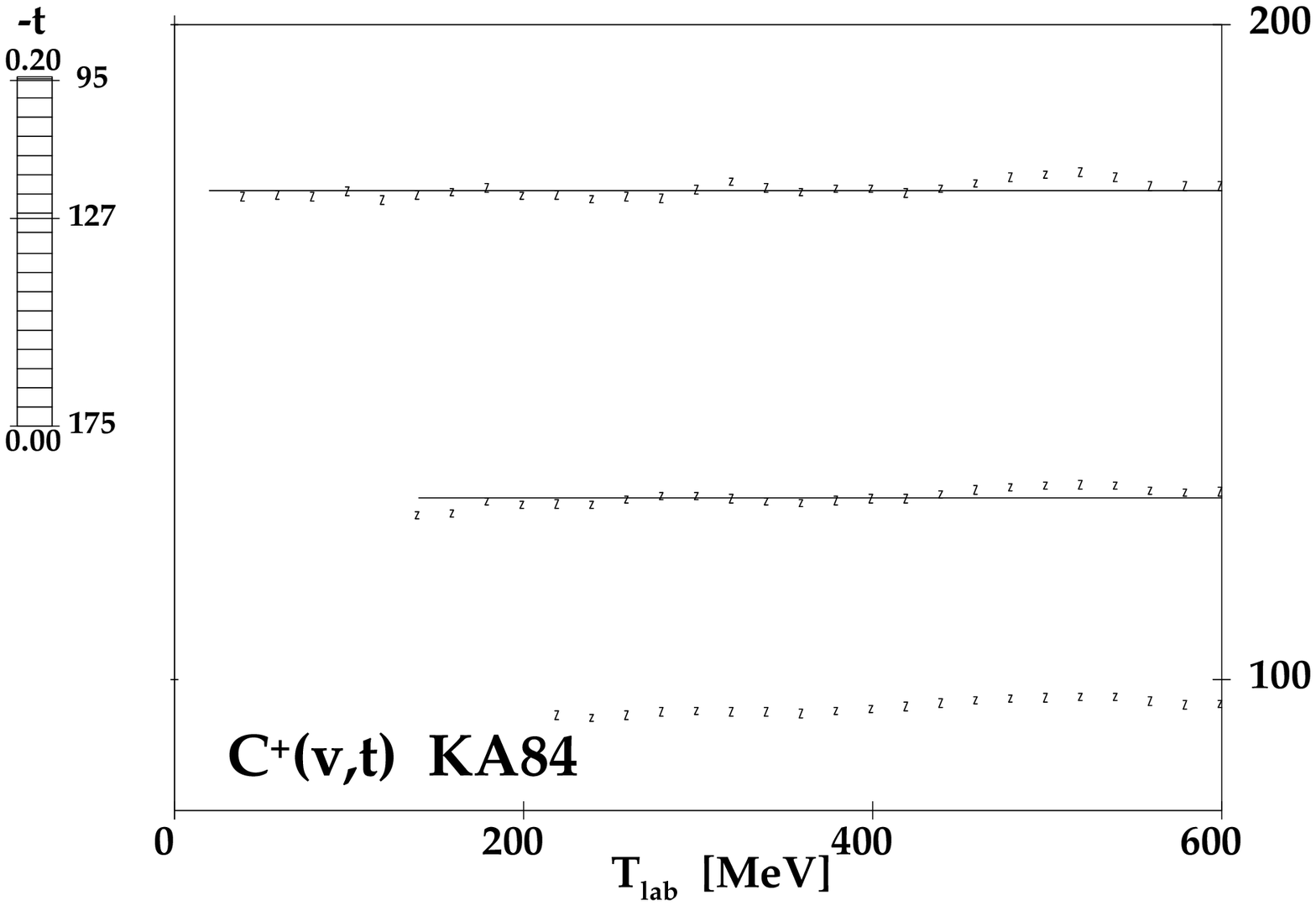, width=0.45\columnwidth}
\end{tabular}
 \par}
  
\caption{\small \setlength{\baselineskip}{2.6ex} \label{fig:c+Fixed-tConst}
  {\bf Left:} Subtraction constants in the fixed-t $C^+$ dispersion
  relation from our $\pi$N analysis SM99 as a function of energy at three
  values of momentum transfer $t$ ; {\bf Right:} same, for the Karlsruhe
  KA84 analysis \protect~\cite{ka84}. This DR yields the data points shown
  in Fig.\protect~\ref{fig:c+Extrap} used to extract the coefficients
  $\bar{d}^{+}_{0i}$ (Eqn.\protect~\ref{eqn:linearSigma}).  }
\end{figure}

The VPI/GW $\pi$N partial-wave and dispersion relation analysis is an
ongoing project, where new solutions are released when changes to the
database and analysis method warrant ~\cite{said}.  Analysis details can be
found in Ref.~\cite{vpi-pwa,pav97,pav99}. Presently, our partial-wave
analysis is constrained by the forward $C^{\pm}(\omega)$ and ``derivative''
$E^\pm(\omega)$ dispersion relations, as well as the fixed-t $B_{\pm}(\nu,
t)$ (``H\"uper" ~\cite{hoehler}) and $C^{\pm}(\nu,t)$ dispersion relations.
These DRs are constrained to be satisfied to within $\sim$1\% up to
$\sim$800 MeV. As our analysis extends up to 2 GeV, the KH80 ~\cite{kh80}
phases are used from 2 to 4 GeV in the dispersion integrals. A 4 GeV cutoff
is sufficient for adequate convergence in the fixed-t $B_\pm$ and $C^{-}$
DR integrals, however the $E^\pm$ and $C^+$ DR integrals require a
parameterization for the high energy parts.  After the report at MENU97
~\cite{pav97}, we included the high energy parts of the latter DRs using
formulas from Ref.~\cite{hoehler}, resulting in much more satisfactory
results.

Pion-nucleon dispersion relations depend on  {\it a priori} unknown
constants including the $\pi$NN coupling constant $f^{2}$ and the
subtraction constants (usually chosen to be scattering lengths).  Our
analysis treats these constants as unknown parameters to be determined by a
best fit to data. In practice, for our work-in-progress
``SM99''~\cite{said}, the coupling $f^2$ and the p-wave scattering volume
$a^{+}_{1+}$ were searched, while the s-wave scattering lengths were taken
from the P.S.I. pionic hydrogen results~\cite{lei98-pc}. We also insisted
that the GMO sum rule ~\cite{gol58} be satisfied.

For every solution, the subthreshold coefficient $\bar{d}^{+}_{00}$ is
calculated using the chosen parameter set and $\pi$N phases from :
 \begin{equation}
   \label{eqn:subtrconst}
   \bar{d}^{+}_{00} = K_{1}\cdot a^{+}_{0+} + 
                K_{2}\cdot f^2 
                + \int d\nu' K_{3}(\nu') Im{D^+(\nu')}
 \end{equation}
 where $K_{i}$ are kinematical factors, and $a^{+}_{0+}$ is the isoscalar
 s-wave scattering length. The expression for $\bar{d}^{+}_{01}$ is
 analogous, involving instead the isoscalar p-wave volume $a^{+}_{1+}$ and
 the amplitudes $E^{+},B^{+},$ and $C^{+}$.  By noting how $\Sigma_d$ varies
 for solutions away from the optimum, and fluctuations of the extracted
 constants with respect to energy, one obtains an indication of the
 uncertainty.  To determine the experimental sigma term $\Sigma$, we use
 $\Delta_D$=12 MeV (see e.g.  Ref.~\cite{sainio97}), which is insensitive
 to the $\pi$N partial wave input ~\cite{glls88,sainio-pc}.
 
 The fixed-t $C^{+}$ DR subtraction constants $C^{+}(\nu=0,t)$ are
 equivalent to $D^{+}(0,t)$.  Thus the slope of these constants as a
 function of $t$ at $t$=0 is $d^{+}_{00} + t\cdot d^{+}_{01}$, so we have
 another method to determine $\Sigma_d$.  Note that these subtraction
 constants are not fixed {\it a priori} in the DR parameter search
 procedure (unlike e.g. $f^2$), so this method of obtaining $\Sigma_d$ is
 independent to the GLLS approach and a valuable consistency check.

\section*{Results and Discussion}

\begin{table}[ht]
  \begin{center}
    \begin{tabular}{|l|r||c|c|c||c|c|c|}
      \hline
      Solution & $\Sigma_{d}$~[MeV] = & ``$a^{+}_{0+}$~const. & Born 
      & $\int$D$^{+}$ & ``$a^{+}_{1+}$''~const. & Born & $\int$E$^{+}$ \\ 
      \hline\hline
      KA84 & 50 = & -7 & +9 & -91 & +352 & -142 & -72 \\
      \hline
      SM99  & 78 = & 1.5 & +9 & -88 & +360 & -136 & -69 \\
      \hline\hline
      {\bf difference} & 28 = & ${\bf +9}$ & {\bf 0} & {\bf +3}
      & ${\bf +8}$ & ${\bf+6}$ & {\bf +3} \\
      \hline\hline
      {\it Expectation} &\it{21$\pm$6}=&\it{8$\pm$3}& 0 &{\it 5$\pm$5} & 0
      & {\it 7$\pm$2} & {\it 2$\pm$4}\\
      \hline
    \end{tabular}
    \caption{Comparison of $\Sigma_d$ from the
      Karlsruhe solution KA84 ~\cite{ka84} and our recent solution SM99
      (values rounded). The change in the $C^+$
      subtraction constant ($a^{+}_{0+}$) term, the $E^+$ Born term, and
      both integral terms are consistent with {\em expectations} from
      pionic atom data \protect~\cite{lei98-pc,loi99}, a lower coupling
      constant ($f^2 \simeq 0.0755$) ~\cite{deSw97}, and a narrower $\Delta$
      resonance width. See text for details.
}
    \label{tab:sigma_terms}
  \end{center}
\end{table}

Our solution ``SM99'' satisfies fixed-t and forward dispersion relations
well (up to our $\sim$800 MeV constraint limit), and the data 
(up to 2 GeV) are
fit with $\chi^{2}$/data point = (2, 2, 2.5) for ($\pi^{+}$,$\pi^-$,CEX).
Compared to the Karlsruhe KA84 solution~\cite{ka84}, these same dispersion
relations are better satisfied (see
Figs.~\ref{fig:c+Const},~\ref{fig:e+Const},~\ref{fig:c+Fixed-tConst}), and
the data much better fit ($\chi^{2}$/point = (4, 5, 3.5) for KA84).  The
PWA and DR solutions clearly favour a $\pi$NN coupling constant
$f^{2}=0.0759\pm0.0004$ ($\frac{g^2}{4\pi}=13.72\pm 0.07$)
\footnote{See our companion article on our $f^2$ determination 
  in these proceedings for details ~\cite{pav99}.}, consistent
with our recently published solutions ~\cite{vpi-pwa}. This value is 
compatible with most recent determinations~\cite{deSw97} and 
$\sim$5\% below the canonical value 0.079 used in the KH80 and KA84
solutions

For the subthreshold coefficients from the GLLS method, we obtain 
$\bar{d}^{+}_{00}$= -1.27$\pm$0.03 
and $\bar{d}^{+}_{01}$= 1.27$\pm$0.03 $m^{-1}_{\pi}$, where
the uncertainty is from the energy fluctuations only
(see Figs.~\ref{fig:c+Const} and ~\ref{fig:e+Const}).  This implies
$\Sigma_{d}\simeq$78 MeV (Eqn.~\ref{eqn:linearSigma}), which is $\sim$55\%
larger than the canonical result $\simeq$50 MeV
~\cite{glls88,sainio97,hoehler}.  As a check of our dispersion relation
machinery, we inputed the Karlsruhe KA84 ~\cite{ka84} phases and reproduced
their $f^{2}$ and $\Sigma_{d}$ results exactly.
Table~\ref{tab:sigma_terms} shows a term by term comparison
between SM99 and KA84 to analyze the differences.

Though the difference between the SM99 and KA84 $\Sigma_d$ values is
surprisingly large, one {\it expects} about 21 MeV of the difference from
new information on pionic atoms, a lower coupling constant, and a narrower
$\Delta$ resonance width.  The isoscalar scattering length $a^{+}_{0+}
\simeq -0.008 ~m^{-1}_{\pi}$ for KA84 (and KH80), but analyses of recent PSI
pionic hydrogen and deuterium results yields $\simeq -0.0015$ ~\cite{loi99}
or $\simeq +0.002$ ~\cite{lei98-pc}. Our analysis used the latter, while the
``expectation'' in Table~\ref{tab:sigma_terms} assumes $0.000\pm0.003$.  A
lower coupling constant around $f^2 =0.0755\pm 0.0010$ is favoured by most
analyses~\cite{deSw97} and this ``expectation'' contributes +7 MeV in
Table~\ref{tab:sigma_terms} from the $E^+$ Born term.  The $C^+$ Born term
does not change due to a well know insensitivity to $f^2$.  And it is well
known that the $\Delta$ resonance width is too wide in KA84 (overshoots the
total cross sections on the left wing), so since $Im{D^+}$ is proportional
to the sum of the $\pi^+$p and $\pi^-$p total cross sections via the
optical theorem, one expects the $D^+$ integral contribution to decrease.
Due to $\Delta$ region dominance of the $C$ DR, the $\Delta$ width and $f^2$
are correlated, and a $\sim 5$\% decrease in $f^2$ {\it roughly}
corresponds to a same decrease in the integrals, and this expectation is
reflected in Table~\ref{tab:sigma_terms}. A narrower $\Delta$ also would
reduce the $E^+$ DR integral, but possible changes in higher partial waves
make predictions less clear. So from rather general
considerations, one {\it expects} a significant increase from the canonical
value for $\Sigma_d$ based on new experimental information.

\begin{figure}
  {\par\centering
  \epsfig{file= 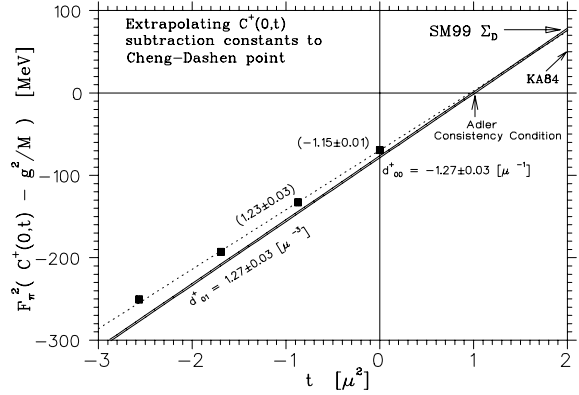, width=0.5\columnwidth}
 \par}
  
\caption{\small \setlength{\baselineskip}{2.6ex} \label{fig:c+Extrap}
Tangent at $t=0$ (dashed line) of the SM99 $\bar{C}^+(0,t)$ 
subtraction constants (solid squares, which include r.m.s. errors),
with tangent inferred from our forward $C+$ and $E^+$ DR analysis overlayed
(solid line).
The slight discprepancy is understood and under investigation. Nonetheless,
 both yield $\Sigma_d \simeq$ 79 MeV, and clearly inconsistent
with the KA84 result $\simeq$ 50 MeV \protect~\cite{hoehler,sainio97}.
}
\end{figure}

The result from the tangent of the $\bar{C}^{+}(0,t)$ subtraction constants
at $t$=0 yields $\bar{d}^+_{00}$= -1.15$\pm$0.03 and
$\bar{d}^+_{01}$=1.23$\pm$0.03, where the uncertainties reflect only the
energy fluctuations of the constants (see Fig.~\ref{fig:c+Fixed-tConst}).
This yields $\Sigma_d$=80 MeV, consistent with our other determination.
Figure~\ref{fig:c+Extrap} shows this result along with the tangent inferred
from the forward $C^+$ and $E^+$ DR analysis. The consistency is not
perfect, and the slight differences in the $\bar{d}^{+}_{0i}$, which are
believed to be understood, are being studied further.

In summary, we have performed a new $\pi$N partial wave and dispersion
relation analysis, from which we obtain $\Sigma$ = 91$\pm$8 MeV using two
different methods, about 27 MeV larger than the canonical result 64$\pm$8
MeV from Ref.~\cite{koch82}. At first glance the result is indeed
surprising, but a large upward change is in fact {\it expected} based on
new information on $a^{+}_{0+}\simeq 0.000$ from pionic hydrogen and
deuterium ~\cite{lei98-pc,loi99}, a lower $\pi$NN coupling constant $f^2
\simeq 0.0755$~\cite{deSw97}, and a narrower $\Delta$ resonance width.
Further study is planned to explore systematic
uncertainties and to resolve small inconsistencies. A new analysis
based on the the Karlsruhe methods~\cite{kh80,ka84} applied to the modern
data is urged to check these findings.

\section*{Acknowledgements}
We gratefully acknowledge a contract from Jefferson Lab under which this
work was done. The Thomas Jefferson National Accelerator Facility
(Jefferson Lab) is operated by the Southeastern Universities Research
Association (SURA) under DOE contract DE-AC05-84ER40150. MMP thanks
A. Badertscher and GWU for their support.

\bibliographystyle{unsrt}

\begin{thebibliography}{99}
\bibitem{gl82} J. Gasser, H. Leutwyler, Phys.\ Rep. \ {\bf 87} 77 (1982)  
\bibitem{glls88} J. Gasser, H. Leutwyler, M.P. Locher, M.E. Sainio.
                  Phys.\ Lett. \ {\bf B213} 85 (1988) ; 
                  {\it ibid} , Phys.\ Lett.\ {\bf B253} 252 (1991) ; 
                  {\it ibid} , Phys.\ Lett.\ {\bf B253} 260 (1991)  
\bibitem{jaffe-pc} R. Jaffe, private communication, 1998.
\bibitem{koch82}  R. Koch, Z.\ Phys.\ {\bf C15} 161 (1982)
\bibitem{hoehler} G. H\"ohler, {\em Pion Nucleon Scattering},
  Landolt-B\"ornstein, Vol.9 b2,   ed. H. Schopper (Springer, Berlin, 1983)
\bibitem{sainio97} M.E. Sainio, 
    {\em Proceedings of 7$^th$ International Symposium on Meson-Nucleon
Physics and the Structure of the Nucleon}, 
(Universit\"at Karlsruhe, UCLA, 1997) eds. D. Dreschsel, G. H\"ohler,
W. Kluge, H. Leutwyler, B.M.K. Nefkens, H.-M. Staudenmaier, 
pp.\ 144--149 ; and references therein.
\bibitem{cheng71} T.P. Cheng and R. Dashen, Phys. Rev. Lett. {\bf 26} 594
  (1971)
\bibitem{kh80}  R. Koch,  E. Pietarinen, Nucl.\ Phys. {\bf A336} 331 (1980)
\bibitem{ka84}  R. Koch, Z.\ Phys.\ {\bf C29} 597 (1984).
\bibitem{hoehler-pc} G. H\"ohler, private communication ; and remarks made
  during the meeting.
\bibitem{said} Recent solutions can be accessed via
 TELNET (or SSH) said.phys.vt.edu (username: {\em said}) , or 
at http://said.phys.vt.edu/ .
\bibitem{vpi-pwa} R.A. Arndt, I.I. Strakovsky, R.L. Workman, M.M. Pavan,
                 Phys.\ Rev.\ {\bf C52} 2120 (1995) ;
 R.A. Arndt, R.L. Workman, M.M. Pavan,
                 Phys.\ Rev.\ {\bf C49} 2729 (1994).  
\bibitem{pav97} M.M. Pavan and R.A. Arndt,
    {\em Proceedings of 7$^th$ International Symposium on Meson-Nucleon
Physics and the Structure of the Nucleon}, 
(Universit\"at Karlsruhe, UCLA, 1997) eds. D. Dreschsel, G. H\"ohler,
W. Kluge, H. Leutwyler, B.M.K. Nefkens, H.-M. Staudenmaier, 
pp.\ 165--169 ; and references therein.
\bibitem{pav99} M.M. Pavan, R.A. Arndt, R.L. Workman, I.I. Strakovsky,
 these proceedings.
\bibitem{lei98-pc} H.J. Leisi, private communication, Oberjoch meeting,
  1998.
\bibitem{gol58} M.L. Goldberger and S.B. Treiman, Phys. Rev. \textbf{110}
  1178 (1958)
\bibitem{sainio-pc} M. Sainio, private communication.
\bibitem{loi99} B. Loiseau, et. al., these proceedings.
\bibitem{deSw97} J.J. deSwart, M.C.M. Rentmeester, and
  R.G.E. Timmermans, 
{\em Proceedings of the 7\( ^{th} \) International Symposium on
Meson-Nucleon Physics and the structure of the Nucleon}
(Universit\"at Karlsruhe, UCLA, 1997) eds. D. Drechsel, G. H\"ohler,
W. Kluge, H. Leutwyler, 
B.M.K. Nefkens, and H.-M. Staudenmaier, page 89.

\end{thebibliography}

\end{document}